# Experimental Implementation of Projective Measurement in Bell Basis

Jae-Seung Lee and A. K. Khitrin

Department of Chemistry, Kent State University, Kent OH, 44242-0001

**Abstract**

A scheme for direct projection of a quantum state on Bell states is described. The method is based on constructing an average Hamiltonian with Bell eigenstates and then, projecting the state on these eigenstates. The projection is performed by adding the results of a direct and time-reversed evolution. Experimental demonstration is shown for pairs of dipolar-coupled nuclear spins.

In quantum mechanics, projective measurement returns one of the eigenvalues of an observed quantity, while the state (wavefunction) collapses to the corresponding eigenstate. Any observable physical quantity can be represented by a Hermitian operator. At the same time, there is no general approach for designing a measurement that projects the state onto eigenstates of an arbitrary Hermitian operator. When it is necessary to perform a projection on the states, which are not eigenstates of some observable, the results of such projection are reconstructed indirectly by applying to the quantum state a set of unitary transformations, before measuring the available observables. This procedure constitutes a basis for the state reconstruction.

Four entangled Bell states of a two-qubit system: $|\Phi_\pm\rangle = 2^{-1/2} (|00\rangle \pm |11\rangle)$ and $|\Psi_\pm\rangle = 2^{-1/2} (|01\rangle \pm |10\rangle)$, played an important role in exploring differences between quantum and classical physics, formulating the Bell inequalities [1], or the EPR paradox [2], discussions on non-locality and hidden variables [3,4]. Discrimination between the Bell states is an important measurement in quantum communication. As an example, the protocols of dense coding [5], quantum teleportation [6], and entanglement swapping [7] require a projective measurement in the Bell basis. Until now, such measurement remained a *gedanken* experiment, and there have been no physical realizations of the direct projective measurement in the Bell basis. Experimental discrimination between the Bell states has been achieved by joint measurements with probabilistic success [8] or by disentangling the Bell states into separable states [9-11]. In this Letter, we describe a direct projection onto the Bell states for a system of two dipolar-coupled nuclear spins.

Nuclear magnetic resonance (NMR) has been an outstanding testbed for implementing controlled dynamics in systems of up to twelve coupled nuclear spins [12-14]. A flexibility of NMR in handling spin dynamics comes from possibility of fast modulation of internal interactions by applying an external radio-frequency field. Such modulation allows "switching" interactions on and off or, more generally, creating average [15] or effective [16] Hamiltonians, which naturally do not exist. The scheme described below is based on creating an average Hamiltonian with non-degenerate eigenvalues and Bell states as its eigenstates. After that, an arbitrary initial quantum state is projected on eigenstates of this average Hamiltonian. The projection, a non-unitary operation, is performed by averaging out the coherences between different eigenstates. Possible NMR schemes, reproducing the results of projective quantum measurement, are described in [17].

The average Hamiltonian with Bell eigenfunctions $H_{Bell}$ can be constructed as follows. Suppose that the eigenvalues corresponding to the eigenvectors $2^{-1/2}$ ($|00\rangle+|11\rangle$), $2^{-1/2}$ ($|00\rangle-|11\rangle$), $2^{-1/2}$ ($|01\rangle+|10\rangle$), and $2^{-1/2}$ ($|01\rangle-|10\rangle$) are, respectively, $a$, $b$, $c$ and $d$. In the multiplicative basis, the Bell Hamiltonian $H_{Bell}$ can be obtained by a unitary transformation

$$H_{Bell} = \frac{1}{\sqrt{2}} \begin{pmatrix} 1 & 1 & 0 & 0 \\ 0 & 0 & 1 & 1 \\ 0 & 0 & 1 & -1 \\ 1 & -1 & 0 & 0 \end{pmatrix} \begin{pmatrix} a & 0 & 0 & 0 \\ 0 & b & 0 & 0 \\ 0 & 0 & c & 0 \\ 0 & 0 & 0 & d \end{pmatrix} \frac{1}{\sqrt{2}} \begin{pmatrix} 1 & 0 & 0 & 1 \\ 1 & 0 & 0 & -1 \\ 0 & 1 & 1 & 0 \\ 0 & 1 & -1 & 0 \end{pmatrix} = \frac{1}{2} \begin{pmatrix} a+b & 0 & 0 & a-b \\ 0 & c+d & c-d & 0 \\ 0 & c-d & c+d & 0 \\ a-b & 0 & 0 & a+b \end{pmatrix}, (1)$$

or, by using the Pauli spin operators,

$H_{Bell} = 4^{-1}$ [ $(a+b+c+d)$ $\mathbf{1}$ + $(a+b-c-d)$ $\sigma_{1z}\sigma_{2z}$ + $(a-b+c-d)$ $\sigma_{1x}\sigma_{2x}$ + $(-a+b+c-d)$ $\sigma_{1y}\sigma_{2y}$ ], (2)

where $\mathbf{1}$ is the identity operator. The terms of this Hamiltonian can be built from the dipole-dipole and the double-quantum Hamiltonians. At $a = -b$ and $d = 0$

$$H_{Bell} = 2^{-1} c \mathbf{1} - 4^{-1} c (\sigma_{1z}\sigma_{2z} - \sigma_{1x}\sigma_{2x} - \sigma_{1y}\sigma_{2y}) + 2^{-1} a (\sigma_{1x}\sigma_{2x} - \sigma_{1y}\sigma_{2y})$$

$$= 2^{-1} c \mathbf{1} - 2^{-1} c H_{zz} + (a/3)(H_{xx} - H_{yy}), \qquad (3)$$

where $H_{zz} = \sigma_{1z}\sigma_{2z} - 2^{-1}(\sigma_{1x}\sigma_{2x} + \sigma_{1y}\sigma_{2y})$ is the secular Hamiltonian of dipole-dipole interaction between spins 1 and 2, $H_{xx} = \sigma_{1x}\sigma_{2x} - 2^{-1}(\sigma_{1y}\sigma_{2y} + \sigma_{1z}\sigma_{2z})$, $H_{yy} = \sigma_{1y}\sigma_{2y} - 2^{-1}(\sigma_{1z}\sigma_{2z} + \sigma_{1x}\sigma_{2x})$, and $H_{xx} - H_{yy}$ is a pure double-quantum Hamiltonian. The Hamiltonian proportional to $H_{xx} - H_{yy}$ can be obtained from $H_{zz}$ by applying the multi-pulse sequence with eight-pulse cycle [18]. $H_{Bell}$ in Eq. (3) has an additional $H_{zz}$ term and, therefore, can be obtained by changing relative intervals between pulses in the eight-pulse sequence. Parameters of the pulse sequence can be optimized by a computer simulation which takes into account difference of chemical shifts for the spins and finite duration of the pulses.

Below we present the results for the case when the initial state is in the subspace spanned by only two of the Bell states $|\Phi_\pm\rangle = 2^{-1/2}(|00\rangle \pm |11\rangle)$. This case is especially simple for experimental realization but well illustrates the principle. The physical system contained 2% of 1-dodecene-1,2-$^{13}C_2$ dissolved in liquid crystal 4'-Pentyl-4-cyanobiphenyl (5CB). Under proton decoupling, $^{13}C$ nuclei of the same molecule form isolated spin pairs with residual dipolar coupling between the spins of the same pair. The experiment has been performed with a Varian Unity/Inova 500 MHz NMR spectrometer at 23°C. At this temperature, the chemical shift difference between two $^{13}C$ spins is 3 kHz. The splitting due to the coupling is 353 Hz, and $^{13}C$ NMR lines from the 1-dodecene-1,2-$^{13}C_2$ molecules and 5CB do not overlap.

The pulse sequence is shown in Fig. 1. The first step, not shown in the figure, is the $^{13}$C polarization enhancement by cross-polarization from protons, performed by two simultaneous frequency-sweeping pulses [19]. During the step A, the pseudopure ground state $|00\rangle$ is created by using a partial saturation. A two-frequency irradiation of 5 ms duration equalizes the populations of three states other than the ground state. Unwanted coherences between states are removed by turning off the $^1$H decoupling [20] for 1 ms after the two-frequency pulse. This elimination of coherences is very efficient, as evidenced by the reconstructed density matrix of the pseudopure ground state in Fig. 2(a). For the state reconstruction we used the protocol described in [21]. When needed, the Bell states $|\Phi_\pm\rangle$ can be obtained from the ground state by using a sequence of rf pulses and delays, as it is shown in step B of Fig. 1. The sequence is $(\pi/2)_h$-$(\tau/2)$-$(\pi)_h$-$(\tau/2)$-$(\pi/2)_s$, where $(\pi/2)_h$ and $(\pi)_h$ are, respectively, non-selective 90 and 180 pulses, $(\pi/2)_s$ is a selective pulse on either of two $^{13}$C spins, and $\tau = 1 / (2 \times 353$ Hz$)$ is the delay producing the phase gate: $|00\rangle \rightarrow |00\rangle$, $|01\rangle \rightarrow |01\rangle$, $|10\rangle \rightarrow |10\rangle$, and $|11\rangle \rightarrow -|11\rangle$. The reconstructed density matrix for one of the Bell states $|\Phi_+\rangle = 2^{-1/2} (|00\rangle + |11\rangle)$ is shown in Fig. 3(a).

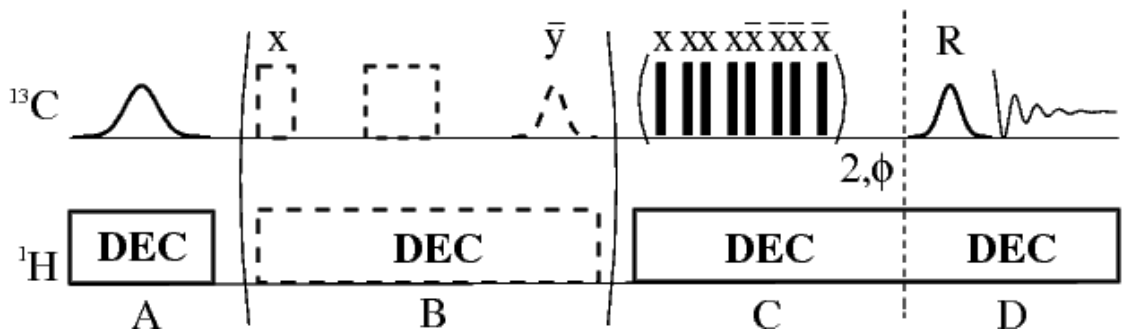

Fig. 1. NMR pulse sequence.

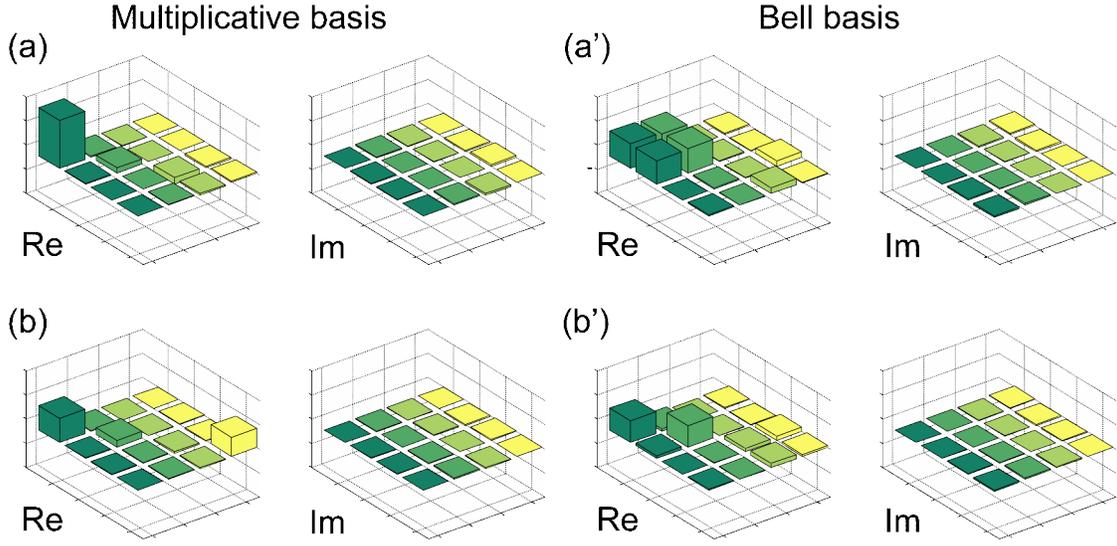

Fig. 2. Reconstructed density matrices for the initial ground state |00⟩ before (a) and after (b) projection on Bell states.

The state projection is performed in step C of Fig. 1. For the subspace of states $|\Phi_\pm\rangle$ one can use the original eight-pulse sequence [18] without any modification. This sequence creates the double-quantum average Hamiltonian with Bell eigenstates. However, two of them $|\Psi_\pm\rangle$ = $2^{-1/2}$ (|01⟩ ± |10⟩) are degenerate and cannot be distinguished. The states $|\Phi_\pm\rangle$ have different eigenvalues $\lambda_1$ and $\lambda_2$. The total duration of two cycles of the eight-pulse sequence $t$ = 1.5 ms has been adjusted to give $t\,(\lambda_1 - \lambda_2) = \pi/2$. The eigenvalues $\lambda_1$ and $\lambda_2$ have been calculated numerically by using experimental values of chemical shifts, coupling constant, and pulse duration. Calculated fidelities of the states $|\Phi_\pm\rangle$ were 0.91 and 1.00. Change of sign of the double-quantum average Hamiltonian, or reversed evolution, can be achieved by shifting the phases of all pulses in the eight-pulse sequence by $\pi/2$. Elimination of the coherences was done by adding the results from the forward and time-reversed evolutions after the evolution time $t = (\pi/2)(\lambda_1 - \lambda_2)^{-1}$, which produced the relative phase $\pi$ between the coherences. As a result, the coherences between two Bell states $|\Phi_\pm\rangle$ have been canceled out and the projection accomplished. In a more general case involving all four Bell states, elimination of coherences can be done with an array of evolution times, similar to used in [12] for averaging out the unwanted coherences.

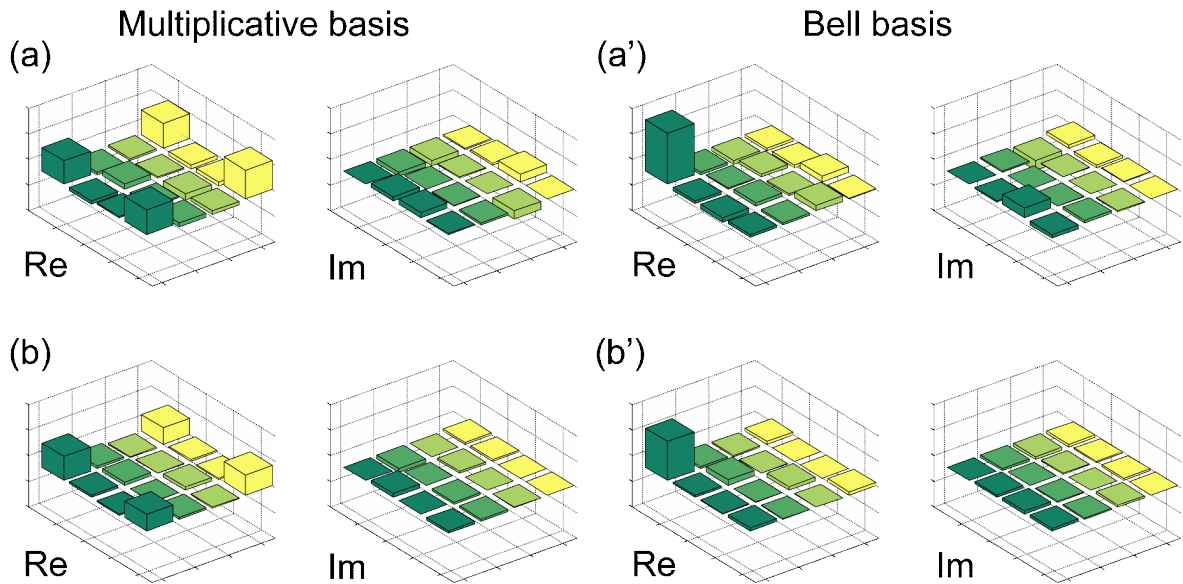

Fig. 3. Reconstructed density matrices for the initial Bell state $|\Phi_+\rangle$ before (a) and after (b) projection on Bell states.

In step D (Fig. 1), a set of the NMR signals was acquired to reconstruct the density matrix [21]. Figs. 2 and 3 show the results for the initial ground state $|00\rangle$ and the Bell state $|\Phi_+\rangle$ before and after the projection. The reconstructed density matrices are presented in both multiplicative and Bell bases. As one can see in Fig. 2(a), the density matrix of the ground state |00⟩ has only one diagonal element in the multiplicative basis. At the same time, this state is a superposition of two Bell states [Fig. 2(a')]. The result of projecting this state onto the Bell basis is shown in Figs. 2(b) and 2(b'). Since the ground state is a superposition of two Bell states: $|00⟩ = 2^{-1/2} (|\Phi_+⟩ + |\Phi_-⟩)$, it is expected that, after the projection, the system would end in the mixed state $2^{-1} (|\Phi_+⟩⟨\Phi_+| + |\Phi_-⟩⟨\Phi_-|)$. The experiment confirms that the system is in the desired mixed state presented in Fig. 2(b'): the density matrix in the Bell basis has two diagonal and no off-diagonal elements. Simultaneously, this state is a mixture of the states |00⟩ and |11⟩, as one can see in Fig. 2(b).

If the state before projection is one of the Bell states, one would expect that projection on the Bell states would leave the state unchanged. The results of testing this are shown in Fig. 3. The experiment is exactly the same as described above, except that the initial state is the Bell state $|\Phi_+⟩$. The density matrix has only one diagonal element in the Bell basis [Fig. 3(a')] and has a superposition form in the multiplicative basis [Figs. 3(a)]. The result of projection shown in Figs. 3(b) and 3(b') confirms that the state is preserved.

In conclusion, average Hamiltonians, resulting from averaging fast-modulated internal interactions, can be designed to have desired eigenstates. A state of a system can be projected on these eigenstates of the average Hamiltonian. As an example, we have demonstrated projection of a state of a two-spin system on its Bell states, which are not eigenstates of any naturally existing Hamiltonian.

This work was supported in part by the NSF under Grant No. ECS-0608846.

References

[1] J. S. Bell, Physics **1**, 195 (1964).

[2] A. Einstein, B. Podolsky, and N. Rosen, Phys. Rev. **47**, 777 (1935).

[3] A. J. Leggett, Found. Phys. **33**, 1469 (2003).

[4] S. Gröblacher, T. Paterek, R. Kaltenbaek, Č. Brukner, M. Żukowski, M. Aspelmeyer, and A. Zeilinger, Nature **446**, 871 (2007).

[5] C. H. Bennett and S. J. Wiesner, Phys. Rev. Lett. **69**, 2881 (1992).

[6] C. H. Bennett, G. Brassard, C. Crépeau, R. Jozsa, A. Peres, and W. K. Wootters, Phys. Rev. Lett. **70**, 1895 (1993).

[7] M. Żukowski, A. Zeilinger, M. A. Horne, and A. K. Ekert, Phys. Rev. Lett. **71**, 4287 (1993).

[8] D. Bouwmeester, J.-W. Pan, K. Mattle, M. Eibl, H. Weinfurter, and A. Zeilinger, Nature **390**, 575 (1997).

[9] M. A. Nielsen, E. Knill, and R. Laflamme, Nature **396**, 52 (1998).

[10] Y.-H. Kim, S. P. Kulik, and Y. Shih, Phys. Rev. Lett. **86**, 1370 (2001).

[11] M. Riebe, *et al.*, Nature **429**, 734 (2004); M. D. Barrett, *et al.*, Nature **429**, 737 (2004).

[12] J.-S. Lee and A. K. Khitrin, J. Chem. Phys. **122**, 041101 (2005).

[13] J.-S. Lee and A. K. Khitrin, Appl. Phys. Lett. **87**, 204109 (2005).

[14] C. Negrevergne, T. S. Mahesh, C. A. Ryan, M. Ditty, F. Cyr-Racine, W. Power, N. Boulant, T. Havel, D. G. Cory, and R. Laflamme, Phys. Rev. Lett. **96**, 170501 (2006).

[15] U. Haeberlen and J. S. Waugh, Phys. Rev. **175**, 453 (1968).

[16] E. B. Feldman, A. K. Hitrin, and B. N. Provotorov, Phys. Lett. A **99**, 114 (1983).

[17] J.-S. Lee and A. K. Khitrin, Appl. Phys. Lett. **89**, 074105 (2006).

[18] J. Baum, M. Munovitz, A. N. Garroway, and A. Pines, J. Chem. Phys. **83**, 2015 (1985).

[19] J.-S. Lee and A. K. Khitrin, J. Chem. Phys, in press.

[20] G. M. Leskowitz, N. Ghaderi, R. A. Olsen, and L. J. Mueller, J. Chem. Phys. **119**, 1643 (2003).

[21] J.-S. Lee, Phys. Lett. A **305**, 349-353 (2002).